\font\bigtitle=cmbx12 at 14pt
\font\title=cmbx12
\font\name=cmr12
\rightline{\name{CERN-TH/96-366}}
\rightline{\name{hep-th/9612191}}
\bigskip
\bigskip
\bigskip

\name

{\centerline{{\bigtitle{Classical Duality from Dimensional Reduction of
Self 
Dual 4-form}}}}
{\centerline{{\bigtitle{Maxwell Theory in 10 dimensions}}}}
\bigskip
\bigskip
\bigskip
{\centerline{\name{David Berman}}\medskip

{\centerline{ Theory Division, CERN, CH 1211, Geneva, Switzerland}}
\smallskip
{\centerline{ Dept. Mathematics, Durham University, Durham, UK}}
\smallskip
{\centerline{ email: David.Berman@CERN.CH}}}
\bigskip
\bigskip
\bigskip
{\centerline{{\title{Abstract}}}
\bigskip

By dimensional reduction
of a self dual p-form
theory on some compact space, we determine the duality generators of the
gauge theory in 4 dimensions. In this picture duality is seen as a
consequence of the
geometry of the compact space. We describe
the dimensional reduction of 10-dimensional self dual 4-form Maxwell
theory to
give a theory in 4-dimensions with scalar, one form and two form fields
that all transform non trivially under duality.  

\vfill
{\name{hep-th/9612191}}

{\name{CERN-TH/96-366}}
 
{\name{21 February}}
\eject

\baselineskip 18pt

{\title{Introduction}}
\bigskip

Recently there has been much interest in field and string theory
dualities. It has been shown that it is possible to explain some gauge
theory dualities from a geometric point of view by considering a self
dual theory living in $M^4 \times K$ where $K$ is some compact manifold. 
When specifying the self duality condition it is necessary to break
general covariance and specify a direction, chosen to be on the compact
space. Different theories from a 4-dimensional point of view then become
related to different directions on the compact space. However, the
physics must be independent of which direction is chosen. Hence the
duality becomes a consequence of general covariance in the self dual
theory.

This elegant explanation of gauge duality has been applied in a
variety of contexts. In several papers [1,2] various authors have shown
how S-duality in Maxwell theory in 4-dimensions can be produced by
considering the compactification of self dual two form theory in
6-dimensions. The rank of the form refers to the gauge potential,
which for a $p+1$ form theory
couples to a p-brane via its world volume. Hence the 6-dimensional theory
concerns a gauge theory coupling to self dual strings of which there has
been much interest recently [3].  The case we will consider in detail is
an extension to the normal ansatz for a self dual 4-form in
10 dimensions compactified on {$T^6$}. We will begin with a brief summary
of the method used to derive the generators of the duality transformation
for Maxwell theory in 4-d from the self dual 6-d theory.  Then we will
discuss the 10-dimensional case and our extension to the usual ansatz.
The resulting theory in 4-d will be derived along with the generators of
the duality group acting on the 4-dimension theory.  The results may be
interpreted in the context of type IIB string theory of which this is a
part of the bosonic sector, where for small values of the string coupling
constant the gravitational sector decouples. Indeed some authors have
used a similar construction to investigate the duality properties of
dyonic p-branes in the context of IIB supergravity [4,5]. We shall not
explore such applications here. 

\bigskip
\bigskip
\bigskip

{\title{S-duality from self dual 6-dimensional Maxwell Theory}}
\bigskip
Here we briefly review the work of Verlinde and Giveon et al [1,2] as
this will
provide the main method for constructing the duality generators in the more
complex case considered later.

It has been shown that determining a covariant action for self dual
p-form fields is nontrivial [6]. Recently, there has been some work
demonstrating how this might be done [7]. However, following [1], we will
describe the
theory in
a non covariant way
via an action for the ordinary Maxwell type theory with the self duality
condition removed and then later impose the self duality constraint as
an extra equation, not derived from the action. 
 The action for a Maxwell
theory in six dimensions is given by: 
$$S=\int_{M^6} dB \wedge H + {1 \over 2} \int_{M^6} H \wedge ^* H,
\eqno(1)$$ where $H\in \Lambda ^3 (M^6)$ and $B \in \Lambda ^2 (M^6)$. We
shall use
the
notation
$\Lambda ^p (M^d) $ for the space of p forms in the d dimensional
manifold, $M^d$. To impose the self
duality relation we use the non covariant equation: 

  $$i_v (H-dB)=0, \eqno (2)$$ where $i_v$ denotes the inner product with
a vector field $v$. 
We then dimensionally reduce the theory; $M^6 \rightarrow M^4 \times T^2$,
including only
the zero modes of fields on $T^2$. The ansatz
for H and B is then as follows: 

 $$ H=F_D a + F b$$ $$ B=A_D a + A b  \eqno (3)$$  where $ F_D, F \in
\Lambda ^2 (M^4) $ and $ A, A_D \in \Lambda ^1 (M^4) $
and
$a,b \in \Lambda ^1 (T^2)$ As will be checked later this ansatz is
consistent
with the self duality in 6-dimensions.  The $a$ and $b$ 1 forms on $T^2$
are the canonical closed 1 forms associated with the non trivial homology
one cycles on the torus. Hence they form a basis for
$H_1(T^2,{\bf{Z}})$. In the canonical basis the intersection matrix is
such that:

 $$\int_{T^2} a \wedge b = 1$$ 

Essentially this means the area of the torus is defined to be 1. The
period
matrix $\Pi$ is given by:

  $$\Pi = \int_{T^2} \left(\matrix{ a \wedge ^* a & a\wedge ^* b \cr
				    b \wedge ^* a & b \wedge ^* b \cr} \right)
        = {1 \over {\tau_2}}\left(\matrix{ 1 & \tau_1 \cr 
	             \tau_1 & |\tau|^2 \cr}\right)    \eqno    (4)  $$ 
First, a direction for the vector field $v$ must be specified so as to
explicitly
impose the self duality condition. We will choose $v$ such that $i_v a =
0$.  Inserting the ansatz (3) into (2) then gives the usual relation for an
abelian field strength;

 $$F=dA  .  \eqno     (5)$$  Later we will repeat the procedure with a
different choice for $v$ and generate the dual theory. The equivalence of
the two
duality related theories is based on the independence of any physics on our
choice of $v$. We then substitute (5)  and (3) into the action (1) and
obtain
a 4-dimensional action after using (4) to do the integration over the compact
space. This gives:

$$S_4 = \int_{M^4} dA \wedge F_D + {1 \over 2} \int_{M^4} ( F_D dA) \Pi {F_D
\choose dA},  \eqno (6)$$ where we have thrown away an irrelevant total
derivative. We are left
with a theory that contains an auxillary field $F_D$ that we may
integrate out of the partition function. The integral is Gaussian in
nature and so the integration is trivial. Of course it is that possible
anomalous effects might appear in the determinant. As such, aspects of
the global topology of the space time will be important. In references
[1,8], the effects of the topology on the path integral are discussed. We
shall not repeat their results here. The equations of motion for $F_D$
give: $$F_D = \tau_2 ^* dA - \tau_1 dA.
$$ Hence $F_D$ becomes identified with the dual field strength and the
self duality
in 6-dimensions of the ansatz (3) can be confirmed. After integrating out
the $F_D$ field we have the usual Maxwell
theory in
4 dimensions with a theta term:

$$S_4 = \int_{M^4} \tau_2 F \wedge ^* F + \tau_1 F \wedge F$$
We can then identify $\tau_1$ and $\tau_2$ with the theta term coupling and the
usual electric charge coupling of the Maxwell theory. (Some overall factors of $2
\pi$ are needed outside the action to be able to make direct comparison with the
usual $\tau$ parameter.)
\smallskip
The generator of the S-duality is found by choosing the other possible direction
for $v$ in the choice of definition of self duality. Taking $v$ such that
$ i_v b=0 $
and then repeating the calculation carried out above gives the following:

$$S_4 = \int_{M^4} {\tau_2 \over |\tau|^2} F \wedge ^* F - {\tau_1 \over
|\tau|^2} F \wedge F$$ Putting $\tau$ together as $\tau=\tau_1 + i \tau_2$
then we have the
familiar S-duality of $ \tau \rightarrow -1/\tau$. The other generator of
SL(2,Z) $\tau \rightarrow \tau +1$ is found by considering the
topological term $F\wedge F$. This has been well explored in the
literature, see for
example reference [8]. Geometrically, $\tau$ and 1 define the basis
vectors for a lattice, L. 
The torus is then described by $C^2/L$. $\tau \rightarrow \tau +1$
and $\tau \rightarrow -1/\tau$ leave the torus invariant. Hence, it is
possible to understand S-duality as a result of the geometry
of some
compactified space. The generators of the duality transformation are found by
different choices of the direction by which self duality is described in the
compactified space. This calculation has been generalized to any
Riemannian Manifold and to
higher dimensional compact spaces by various authors [2], including the
case of $K^3 \times T^2$, which would be of particular interest from a
string theory point of view. Dimensional reduction of self dual p-form
gauge theories has been investigated recently in other contexts by [9,10].

\bigskip
\bigskip

 {\title{Theories in higher dimensions}}
 \bigskip

Crucial to the discussion above was the existence of a self dual 2 form
gauge
theory in 6 dimensions. As is well known [11], self dual (chiral) gauge
theories
only exist
when the dimension of spacetime D= 2 mod 4. Hence for the next case we
will
look at D=10 self dual 4 form abelian gauge theory. This will also
provide us
with a far richer duality group since the compact manifold will be 6
dimensional. \smallskip First we will deal with the most immediate
generalization of the above
calculation. It is useful to consider the general context of how the
p-form field strength is decomposed between the compact, D dimensional
space, $K^D$ and the, d dimensional space-time, $N^d$. Hence, in general

$$H \in \Lambda^{(D+d)/2} (M^{D+d}) $$
Set $p=(D+d)/2$
$$\Lambda^p (M^{D+d}) = \Lambda^{d/2} (N^d) \otimes \Lambda^{D/2} (K^D)
\bigoplus_i
[\Lambda ^i (N^d) \otimes \Lambda^{p -i} (K^D) \oplus \Lambda^{p - i}
(N^d)
\otimes \Lambda^i (K^D)]$$ Where $i$ runs from 0 to $i \le d$ and $i \le
D$
but not $i$ such that $p-i =
D/2$.  The space has been split up according to the action of the Hodge
operation on elements in the space. The first space in this decomposition
is singled out because the Hodge operation maps an element in the space
back into the same space. The other spaces in the direct sum appear in
pairs with the Hodge operation mapping an element in one space on to the
paired space. For self duality to be possible the field strength of the
theory must be
an element of a space that is mapped back into the same space by the
Hodge operation. The usual ansatz is to truncate the theory, taking fields
that live only in the first space. This was the case with the example
given above and in the ansatz used in reference [4,5]. Fields that are in
this space may couple to dyonic objets and hence are natural candidates
for electromagnetic duality.

If we consider the sum of an element of one
space with an element of its
pair, then we can see that the Hodge operation will map such a sum of
elements back onto the same space. That is the space formed by the direct
sum of a space with its pair. So it is possible to have electromagnetic
self duality for fields living in such a pair of spaces. Later we shall
develop this idea showing using a suitable ansatz, that it is possible to
form such pairs that transform non trivially under a duality transformation.

We shall begin our analysis with fields in the first space. It is
necessary to
have a basis of $H^{(d/2)}(K^d,{\bf{Z}})$ Let $\lbrace \gamma_I \rbrace$
be such a basis. $I=1$
to $b^{(d/2)} (K^d)$ where $b^{(d/2)} (K^d)$ is the $d/2$ Betti number
associated
with the compact manifold $M^d$. A canonical basis is chosen such that the
intersection matrix $$Q_{IJ}=\int_{K^d} \gamma_I \wedge \gamma_J$$ is
antidiagonal.  The period matrix is: $$G_{IJ}=\int_{K^d} \gamma_I \wedge
^*\gamma_J $$ For definiteness we shall now go to the case of $M^{10}$
compactified on
$M^4 \times T^6$. So, the $\lbrace \gamma_I \rbrace$ are the basis of
3-forms in $T^6$ and
$b^3(T^6)=20$. As it is a torus, the three form basis may be written in
terms of a product of the one form basis.
The action in 10 dimension is:
$$S_{10}=\int_{M^{10}} dC \wedge H + {1 \over 2} \int_{M^{10}} H \wedge ^*
H   
\eqno (7)$$
with also the self duality equation:
$$ i_v(H-dC)=0   \eqno (8)$$ The ansatz taken will be as follows:
$$H=\sum_{I=1}^{20} F^I
\gamma_I$$ $$C=\sum_{I=1}^{20} A^I \gamma_I    \eqno (9)$$ where $F^I \in
\Lambda ^2 (M^4)$ and $A^I \in \Lambda ^1 (M^4)$ We then proceed as
before. Picking out a particular $v$ and then decomposing
the 3-form basis, $\lbrace \gamma_I \rbrace$ into parallel and
perpendicular parts via the
equations:  $ i_v \gamma_a \ne 0$ for $\gamma_a$ in the space
{\it{parallel}} to $v$ and $i_v \gamma_i =0 $ for $\gamma_i$ in the space
{\it{perpendicular}} to $v$. From now on the indices a,b will indicate
$\gamma$ parallel and i,j will indicate $\gamma$ perpendicular. This
projection decomposes the 3 form basis into 10 parallel and 10
perpendicular basis 3-forms. Hence, when we substitute the ansatz (9) into
the self duality equation (8) we arrive at 10 equations:$$ F^a=dA^a. 
\eqno(10)$$ First compactify $M^{10}$ on $M^4 \times T^6$ which involves
substituting
in the ansatz (9) into (7) and then doing the necessary integrals over
$T^6$, (as before keeping only the zero modes).  We arrive at a
4-dimensional action. We then substitute in the equations (10) derived
from the self duality equation (8) for a particular choice of $v$. Giving
the action:  $$S_4=\int_{M^4} dA^a \wedge F^i Q_{ai} + {1 \over 2}
\int_{M^4} d A^a \wedge ^*dA^b G_{ab} + 2 F^i \wedge ^*dA^a G_{ia} + F^i
\wedge ^* F^j G_{ij} \eqno(11)$$

As before we must now integrate out the auxilary fields, $F^i$. Again,
the integrals will be Gaussian. Doing the integration gives the following 
action for an abelian gauge theory:
$$S_4=\int dA^a \wedge dA^b \sigma_{ab} + dA^a \wedge ^* dA^b \tau_{ab}
\eqno(12)$$
Where the coupling matrices $\tau$ and $\sigma$ are given by:
 $$\tau_{ab} = G_{ab} + Q_{ai}G^{ij}Q_{jb} - G_{ai}G^{ij}G_{jb}$$
 $$\sigma_{ab}= Q_{ai} G^{ij} G_{ib} + G_{ai} G^{ij} Q_{jb}  \eqno(13)$$ 
The raised indices indicate the inverse matrix, such that
$G^{ab}=G_{ab}^{-1}$ and not the {\it{parallel}} components of
$G_{IJ}^{-1}$.  Note that apart from the usual curvature squared term
there is also a topological term that is a generalization of the theta
term for a U(1) gauge theory. We will see how the above transforms under
duality when we calculate the more general case discussed below.

Before, we noted that it may be possible to construct theories that are self
dual that contain fields of different form rank (We mean the rank of the
fields after compactification; before compactification it is clear that the
fields must have the same rank). These fields will live in
the sum of paired spaces discussed earlier. As such, the ansatz considered
previously may be viewed as a degenerate case; in which the pair of the
space is the space itself. We now move on to consider the case where we
have a sum of fields that live
in such a pair of spaces. Replace the ansatz (9) with the following:
$$H=\sum_I A^I \mu_I  + B^I \nu_I$$
$$C=\sum_I a^I \mu_I  + b^I \nu_I        \eqno (14)$$ where $$A^I \in
\Lambda^1 (M^4) \quad B^I \in \Lambda ^3 (M^4) \quad a \in
\Lambda^0 (M^4) \quad b \in \Lambda^2 (M^4)$$ and $\lbrace \mu_I \rbrace$
is the canonical basis of $H^4(T^6,{\bf{Z}})$
and $\lbrace \nu_I \rbrace$ is the
canonical basis of $H^2(T^6,{\bf{Z}})$. Note $b^2(T^6)=b^4(T^6)=15$. 
Hence, I=1..15. As before we construct the period matrices associated
with both bases. Let G be the period matrix of the 4-form basis,
$\lbrace \mu_I \rbrace$ and
F be the period matrix of the 2-form basis $\lbrace \nu_I \rbrace$. There
will also be an
intersection matrix Q defined by: $Q_{IJ} = \int_{T^6} \mu_I \wedge \nu_J$
which in the canonical basis will be antidiagonal.  We now define
{\it{parallel}} and {\it{perpendicular}} bases for both the 2 and 4 forms
as before. The indices a,b indicate parallel 2-form and i,j indicate
perpendicular 2-form. $\tilde{a},\tilde{b}$ denotes parallel 4-form and
$\tilde{i},\tilde{j}$ denotes perpendicular 4-forms.  It can be seen for
any given one form there are 5 parallel and 10 perpendicular 2-forms
and 10 parallel, 5 perpendicular 4-forms. Hence substituting in the ansatz
(14) into the self duality equation (8) we have for a particular choice
of $v$ a set 15 equations:
$$A^{\tilde{a}}-da^{\tilde{a}}=0$$
$$B^a-db^a=0. \eqno (15)$$ We compactify $S_{M^{10}}$ as before,
performing the necessary $T^6$
integrals which introduce the period and intersection matrices defined
above. Then substitute in the self duality equations (15) into the
compactified action. After throwing away irrelevant total derivatives, we
integrate out all auxillary fields, $B^i$ and $A^{\tilde{i}}$. This
leaves, the following 4-dimensional action:

  $$S_4= \int_{M^4}{1 \over 2}[
da^{\tilde{a}} \wedge ^* da^{\tilde{b}} \tilde{\tau}_{\tilde{a}\tilde{b}}
+ db^a \wedge ^*db^b \tau_{ab} ] - da^{\tilde{a}} \wedge db^b
\sigma_{\tilde{a} b}, \eqno (16)$$ where we have the following couplings: 
$$\tilde{\tau}_{\tilde{a}\tilde{b}} = G_{\tilde{a}\tilde{b}} +
Q_{\tilde{a} i} F^{ij} Q_{j \tilde{b}} - G_{\tilde{a} \tilde{i}}
G^{\tilde{i} \tilde{j}} G_{\tilde{j} \tilde{b}} $$
$$\tau_{ab}=F_{ab} + Q_{a \tilde{i}} G^{\tilde{i} \tilde{j}} Q_{\tilde{j}
b} - F_{ai} F^{ij} F_{jb}$$
$$\sigma_{\tilde{a} b} = Q_{\tilde{a}i} F^{ij} F_{jb} + G_{\tilde{a}
\tilde{i}} G^{\tilde{i} \tilde{j}} Q_{\tilde{j} b}.     \eqno(17)$$ The
action contains the usual kinetic terms for scalar fields and
2nd rank tensor fields. There is also the topological term which is a
generalization of the theta term that couples the scalar and two form
fields.

Note that if $G=F$ then we have the same equation for the coupling as
before, (13). This confirms that the previous case is a
degenerate version of the more general situation in which we have a
paired space. Also, one can easily check
that this formula for the
coupling of the abelian gauge theory in terms of the period matrix of the
compactified space reproduces the simple $T^2$ result. In this instance
the period matrix is two dimensional and so the perpendicular and parallel
parts are one dimensional and hence no matrix inverses are involved.

Now we wish to construct the generators for the duality transformation for
the
theory described above. We follow the previous S-duality example, by
choosing different directions for $v$ and then determining how the
theory changes. It is obvious from equations (16) and (17) that only the
coupling matrices change when a different direction for $v$ is choosen.
Hence, the duality related theories will have the same form with only the
couplings being different. This implies ofcourse that the equations of
motion of the duality related theories will only differ by the value of
the coupling matrices given in the action. The equations of motion for
the action (16) are simply the free field equations for scalar, one form
and two form fields. Also each field strength has the usual Bianchi 
identity. The topological coupling between the scalar and two
form fields will be transparent to the classical equations of motion
but will play a role in the partition function (by analogy with the
usual theta term).

To simplify the calculation we calculate the coupling matrices for a
compact space
that is a product of three orthogonal generic 2-tori, each with area one.
Obviously, $b^1(T^6)=6$, so there are six possible choices for v. Each
choice will give a duality related theory.

The coupling constant matrices were calculated explicitly for each choice
of $v$. These gave:

For $v=a1$ (corressponding to the $a$ cycle of the 1st torus)

$$\tau={1
\over{(^1\tau_{22})}}diag({}^1\tau_{22},{}^2\tau_{11},{}^2\tau_{22},
{}^3\tau_{11},{}^3\tau_{22}) \eqno(18.a)$$ $$\tilde{\tau}=1 \oplus
\left(\matrix{^2\tau_{11} {}^3\tau_{11} & {}^2\tau_{12} {}^3\tau_{12} \cr
{}^2\tau_{12} {}^3\tau_{12} & {}^2\tau_{22} {}^3\tau_{22} \cr
}\right)\oplus \left(\matrix{{}^2\tau_{11} {}^3\tau_{22} & {}^2\tau_{12}
{}^3\tau_{12} \cr {}^2\tau_{12} {}^3\tau_{12} & {}^2\tau_{22}
{}^3\tau_{11} \cr }\right) \oplus 1 \oplus {1
\over{({}^1\tau_{22})}}diag({}^2\tau_{11},{}^2\tau_{22},
{}^3\tau_{11},{}^3\tau_{22}) \eqno(18.b)$$ $$\sigma = {{}^1 \tau_{12}
\over {{}^1 \tau_{22}}} M \eqno(18.c)$$

The direct sum refers to blockdiagonal decomposition of the matrices.  M
is a $5 \times 10$ matrix which has a $4 \times 4$ identity matrix in the
last block and zeros elsewhere. $^a\tau_{ij}$ refers to the ijth
element of the period matrix of the a th 2-torus. This essentially shows
how the moduli of the tori combine to give the coupling constants for the
4 dimensional theory.
\medskip
The coupling matrices calculated for different choices of v can be
related to the
above coupling matrices (calculated for $v$ lying in the $a1$ direction),
as
follows: 

For $v=b1$ (corressponding to the $b$ cycle of the first torus)

$${}^1\tau_{22} \leftrightarrow {}^1\tau_{11} \eqno(19.1)$$

For $v=a2$ (corressponding to the $a$ cycle of the second torus)

$${}^2\tau_{ij} \leftrightarrow {}^1\tau_{ij} \eqno(19.2)$$

For $v=b2$

$${}^1\tau_{11} \leftrightarrow {}^2\tau_{22}$$

$${}^1\tau_{12} \leftrightarrow {}^2\tau_{12} \eqno(19.3)$$

For $v=a3$

$${}^1\tau_{ij} \leftrightarrow {}^3\tau_{ij} \eqno(19.4)$$

For $v=b3$

$${}^1\tau_{11} \leftrightarrow {}^3\tau_{22} $$

$${}^1\tau_{12} \leftrightarrow {}^3\tau_{12} \eqno(19.5)$$

These form a
set of duality generators that act on the couplings of the
theory. Note, the transformation $\tau_{11} \leftrightarrow \tau_{22}$ is
equivalent to the imaginary part of $ \tau \leftarrow {-1 \over \tau}$. So
the duality generators we have above correspond to a generalization of the
coupling inversion duality generator of SL(2,Z).

The difference between the S-duality (coupling constant inversion) and
the transformations we describe above is that part of the coupling
constant matrix is left invariant by the duality transformation. In the
terms of our scheme for calculating these generators, this is a result of
having cycles cycles that contain both projection directions. These are
left invariant by the duality transformation; cycles that contain neither
are of course projected out and so do not appear. The cycles that contain
one of the directions are those that are transformed under duality.
Simple counting of the number of 2 and 4 cycles with these properties
confirms this picture.

To determine the other
generators it is neccessary to look at the topological coupling, $\sigma$
and determine the generators that corresponds to the SL(2,Z), $\tau
\rightarrow \tau +1$ Following the same arguements that lead the theta
term being invariant in
the partition function under $\tau \rightarrow \tau + m$ (where m is
integer), we conclude that $\sigma \rightarrow \sigma + m$ leaves the
partition function invariant. In terms of $\tau$ this is equivalent to
$${}^1 \tau_{12} \rightarrow ^1\tau_{12} + m ^1\tau_{22} \eqno(20.1)$$
(for $v$ chosen
to be the $a$ direction of the 1st torus) and $${}^1 \tau_{12} \rightarrow
^1\tau_{12} + m
^1\tau_{11} \eqno(20.2)$$ (for $v$ chosen to be in the $b$ direction of
the
1st torus).

The generators are obviously given by taking $m=1$. Now, to calculate the
group it is neccessary to determine how combinations of the generators
act. Indeed it is clear that the set of generators presented here is not
the minimal set. For example, (19.3) can be formed by composing (19.1) and
(19.2). The minimal set will be given by (19.1), (19.2), (19.4) and
(20.1). Examining the compositions of these generators we find that we
find that they generate the group $SL(2,Z) \times SL(2,Z) \times SL(2,Z)$.
Such a group is what would be naively expected given that the compact
space was taken to be a product of three orthogonal (area=1) tori. For the
most
general 6-torus, the explicit calculation of the
coupling matrices in terms of the modular parameters of the
6-torus would be far more complicated. For such a case we would expect the
group to be the modular group of the six torus with a non linear
realization in terms of the coupling matrices. What we wish to stress here
is the possible introduction of fields that live in the {\it{paired}} 
space, scalar and antisymmetric fields, that
have couplings that tranform under some duality group associated with
a compact space.
   
\bigskip
\bigskip
\bigskip
{\title{Conclusions}}
\bigskip

We have investigated some aspects of duality induced by dimensional
reduction of a self dual p-form theory. It has been shown that a $T^6$
dimensional reduction of a self dual 4-form may include scalar and
antisymmetric
tensor fields as well as 1-form fields. The coupling matrices for these
theories have been calculated. It was shown that the form of the theory is
left invariant but the couplings transform 
non trivially under duality. Indeed, we express the coupling matrices
as functions of
the moduli for the torus. The generators of the duality transformation 
have been explicitly calculated and the group formed by composing these
generators is described, giving the expected modular group. The
possibility of including scalar and
antisymmetric tensor fields may have
interest when considering massive gauge theories in 4-dimensions. 

\bigskip
\bigskip
{\title{Acknowledgements}}\bigskip
I would like to thank David Fairlie for helpful discussions and valuable
comments and the referee for pointing out ref[5]. This
work was supported by PPARC.
\bigskip
\bigskip

{\title{References}}}
\bigskip

[1] E.Verlinde, Nuc. Phys. {\bf{B455}} (1995) 211
 
[2] A.Giveon and M.Porrati, Phys. Lett. {\bf{B385}} (1996) 81
 
[3] R. Dijkgraaf, E. Verlinde, H. Verlinde, Nucl.Phys.{\bf{B486}} (1997)
89
 
[4] M.B.Green, N.D.Lambert, G.Papadopoulus and P.K.Townsend,

Phys. Lett. {\bf{B384}} (1996);
 
[5] Bergshoeff, Boonstra and Ortin, Phys.Rev. {\bf{D53}} (1996) 7206

[6] N.Marcus and J.Schwarz, Phys. Lett. {\bf{B115}} (1982) 111

[7] P.Pasti, D.Sorkin and M.Tonin, Phys. Rev. {\bf{D52}} (1995) 4277
    
    P.S.Howe, E.Sezgin and P.C.West hep-th/9702008 and hep-th/9702111
    
    N.Berkovits, Phys. Lett. {\bf{B388}} (1996) 743

[8] E.Witten, hep-th /9505186
 
[9] I.Giannakis and V.P.Nair, hep-th /9702024

[10] S.Deser, A.Gomberoff, M.Henneaux and C.Teitelboim, hep-th /9702184

[11] M.Henneaux and C.Teitelboim, Phys. Lett. {\bf{B206}} (1988) 650

 \end